# Temperature dependence of self-trapped exciton luminescence in nanostructured hafnia powder

A.O. Shilov[1], S.S. Savchenko[1], A.S. Vokhmintsev[1],
V.A. Gritsenko [2] and I.A. Weinstein [1,a]

[1]NANOTECH Centre, Ural Federal University,
Mira str., 19, Ekaterinburg, Russia, 620002

[2]Rzhanov Institute of Semiconductor Physics SB RAS,
13 Lavrentiev Ave., Novosibirsk, Russia, 630090

a) Corresponding author: i.a.weinstein@urfu.ru

## Abstract

The intrinsic optical properties and peculiarities of the energy structure of hafnium dioxide largely determine the prospects for applying the latter in new generation devices of optoelectronics and nanoelectronics. In this work, we have studied the diffuse reflectance spectra at room temperature for a nominally pure nanostructured $HfO_2$ powder with a monoclinic crystal structure and, as well its photoluminescence in the temperature range of 40 – 300 K. We have also estimated the bandgap $\underline{E_g}$ under the assumption made for indirect (5.31 eV) and direct (5.61 eV) allowed transitions. We have detected emission with a 4.2 eV maximum at T < 200 K and conducted an analysis of the experimental dependencies to evaluate the activation energies of thermal quenching (140 meV) and enhancement (3 meV) processes. Accounting for both the temperature behavior of the spectral characteristics and the estimation of the Huang-Rhys factor $S \gg 1$ has shown that radiative decay of self-trapped excitons forms the mechanism of the indicated emission. In this case, the localization is mainly due to the interaction of holes with active vibrational modes of oxygen atoms in non-equivalent ($O_{3f}$ and $O_{4f}$) crystal positions. Thorough study of the discussed excitonic effects can advance development of hafnia-based structures with a controlled optical response.

## Highlights

- The diffuse reflectance spectra of hafnia nanopowder are investigated
- Intrinsic absorption edge is formed by indirect and direct allowed transitions
- Temperature effects in the 4.2 eV luminescence of self-trapped exciton are analyzed
- Interaction of STE with vibrational modes of non-equivalent oxygens are discussed

# 1. INTRODUCTION

Currently, the use of solid-state $HfO_2$-based structures has good prospects for production of radiation-resistant and optically active coatings, for replacement of $SiO_2$ in CMOS transistor gates, and for creation of hardware components for new generation nanoelectronics [1–8]. Due to its high density and high atomic mass, nanostructured hafnium dioxide is also necessary in designing up-to-date scintillation media doped with various rare earth ions [9,10]. In practice, the high-tech fields mentioned above require a $HfO_2$-based material should be comprehensively investigated to gain a deeper insight into the features of its electronic structure, the regularities of forming optical and luminescent properties, as well as the fundamental mechanisms of the transfer and relaxation of excitations of different nature through intrinsic and impurity defects underlying active centers.

It is well known that hafnium dioxide can be in three different phases at atmospheric pressure: a monoclinic (m-$HfO_2$) phase that converts into a tetragonal one, when heated above 1700 K, and a cubic one at temperatures > 2500 K [11,12]. It is worth emphasizing that the tetragonal and cubic phases in pure $HfO_2$ cannot be preserved at room temperature even by utilizing ultrafast quenching [12]. Thus, the monoclinic-phase structures as the most stable and the most common ones under atmospheric conditions are of greatest interest for research. From the point of view of the experimental study of perfect simulation objects, only one work is today known for the authors to judge synthesized and investigated optical properties of monoclinic single crystal hafnium dioxide [13]. Meanwhile, commercial high-purity-degree powders (99.9%) are often used for analyzing physical and chemical properties [14–16]. Moreover, there are widespread methods to synthesize $HfO_2$ thin films supported on quartz and silicon substrates by atomic layer deposition and magnetron sputtering [9,17–21]. However, the low-dimensional structures obtained in such a way are, as a rule, amorphous and crystallize only after high-temperature annealing [9,22].

Despite the availability of samples with different morphologies to test, the literature chiefly comprises data on luminescence for $HfO_2$ films doped with ions of various chemical elements [9,12,17,21–25]. In this context, the properties of luminescent nominally pure structures still remain largely unexplored. It is known that, when exposed to UV radiation, hafnium dioxide exhibits a 2.7 eV intense blue luminescence associated with oxygen vacancies [3,25,26]. In [27], the possibility is shown to employ this emission in hafnia doped with trivalent europium for development of a potential white LED phosphor.

$HfO_2$ with a monoclinic crystal lattice and structures in the amorphous state are materials possessing a wide energy gap ($E_g$ > 5.5 eV) [9,17,20,23,28,29]. When excited by photons with energies $hv > E_g$, some of among those showcase 4.0 – 4.5 eV emission caused by radiative recombination processes assisted by self-trapped excitons (STE) [9]. In particular, the existence of STE in $HfO_2$ thin films is indicated by a Stokes shift of about 1.5 eV and a strong temperature dependence of the parameters of this luminescence

[10,30]. The present paper elaborately examines the behavior of the STE photoluminescence spectra in the nanostructured nominally pure hafnium dioxide powder over a wide temperature range and estimates appropriate fundamental parameters of the exciton-phonon interaction.

## 2. EXPERIMENTAL

In this work crystalline hafnium dioxide powder (HFO-1 grade, TU 48-4-201-72) was investigated. According to the data of the mass-spectral analysis provided by the manufacturer the powder contains 99.9% $HfO_2$, the Zr impurity does not exceed 0.1% and the other impurities are less than 0.01%.

The images of the powder were obtained using scanning electron microscope (SEM) Carl Zeiss SIGMA VP with Oxford Instruments X-Max 80 module for energy dispersive analysis to determine chemical composition of the sample. X-ray diffraction (XRD) was analyzed using Rigaku Corporation MiniFlex 600 diffractometer with CuK$\alpha$ radiation and SmartLab Studio II software. XRD-patterns were measured for the $2\theta$ range from 5° to 105° with 0.02° step. Raman spectrum was registered by Renishaw U1000 spectrometer in range 50 – 850 $cm^{-1}$. Cobolt Samba green solid-state laser with wavelength of 532 nm and a power of 5mW was used as an excitation source.

Diffuse reflectance spectra (DRS) were registered in the 210-850 nm range with the step of 0.1 nm using two-beam spectrophotometer SHIMADZU UV-2450 and integrating sphere ISR-2200 attachment. Barium sulfate powder was used as a white body reference. The investigations of the photoluminescent (PL) properties of the hafnia powder were carried out using Perkin Elmer LS55 spectrometer. The emission was registered with 50 μs delay after excitation, the data was being integrated for 12.5 ms. The scan speed was set 300 nm/min. The spectral slit of the monochromator was 10 and 5 nm in the excitation and the emission channels, respectively. A xenon lamp was used as a source of photoexciting. To study the temperature effect on the PL spectra the sample was placed into Janis CCS 100/204N closed cycle refrigerator coupled with LakeShore DT-670B-CU temperature sensor and Model 335 controller.

## 3. RESULTS

The SEM images of the investigated powder are shown in Fig. 1. According to the data obtained (see Fig. 1a and b) the powder consists of individual grains with a size of 10 – 400 μm. When zoomed in, see Fig. 1c, the grains are found to be nanostructured, because they are composed of the smaller particles of 40 – 220 nm. Distribution histogram *f(d)* for nanoparticles' size is shown in Figure 1d. The resulting dependence can be described by a lognormal law with a median size 67 nm and the standard deviation 15.5 nm, see the red curve in Fig. 1d. Analysis of the powder chemical composition revealed the presence of 32 ± 2 at.% hafnium and 68 ± 2 at.% oxygen, the resulting ratio, taking into account the error, is close to stoichiometric. Impurities of heavy elements have not been identified.

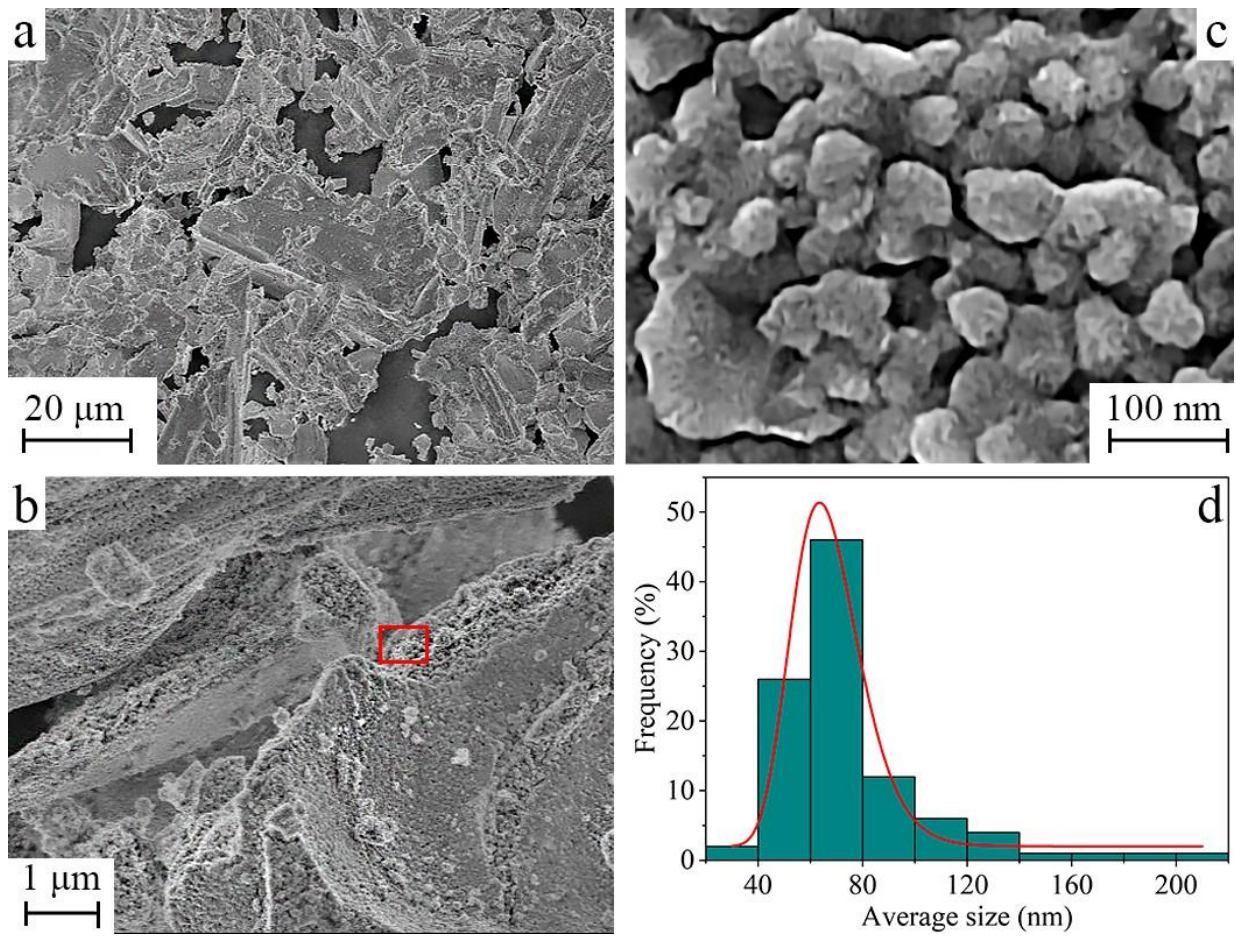

**FIGURE 1.** SEM images of the HfO$_2$ powder studied at different scales (a, b, c) and a histogram of the nanoparticle size distribution (d).

In Fig. 2 the XRD data, Raman and the diffuse reflectance spectra are presented. The XRD pattern in Fig. 2a completely matches monoclinic HfO$_2$, P2$_1$/c space group [33]. The unit cell consists of 4 hafnium atoms and 8 oxygen atoms: 4 atoms with coordination number 3 (O$_{3f}$) and other with coordination number 4 (O$_{4f}$). The diffractogram was used to evaluate the coherently scattering domain size by Williamson-Hall method [31,32]. The calculated value of 68 nm is in a good agreement with the median size of the lognormal distribution, see Fig.1d.

Raman spectrum demonstrates 18 active vibrational modes (9$A_g$ + 9$B_g$) typical for m-HfO$_2$, see Fig. 2b. Raman modes in range of $\omega < 200$ cm$^{-1}$ are related to the hafnium vibrations, while the vibrations in range of $\omega > 270$ cm$^{-1}$ are assigned to the oxygen atoms. The $A_g$ mode with $\omega = 499$ cm$^{-1}$ is characterized by the stretching vibration of the shortest Hf–O bond, at which O$_{3f}$ atoms move closer to its nearest neighbors while O$_{4f}$ show weak vibrations [34].

According to Fig. 2c it should be mentioned that the reflectance is exceeding 95% in near infrared and visible spectral ranges. For $\lambda < 400$ nm slight decrease is observed with local minimum near 250 nm. In the $\lambda < 240$ nm region sharp decrease is observed, which can be caused by optical transitions near intrinsic absorption edge.

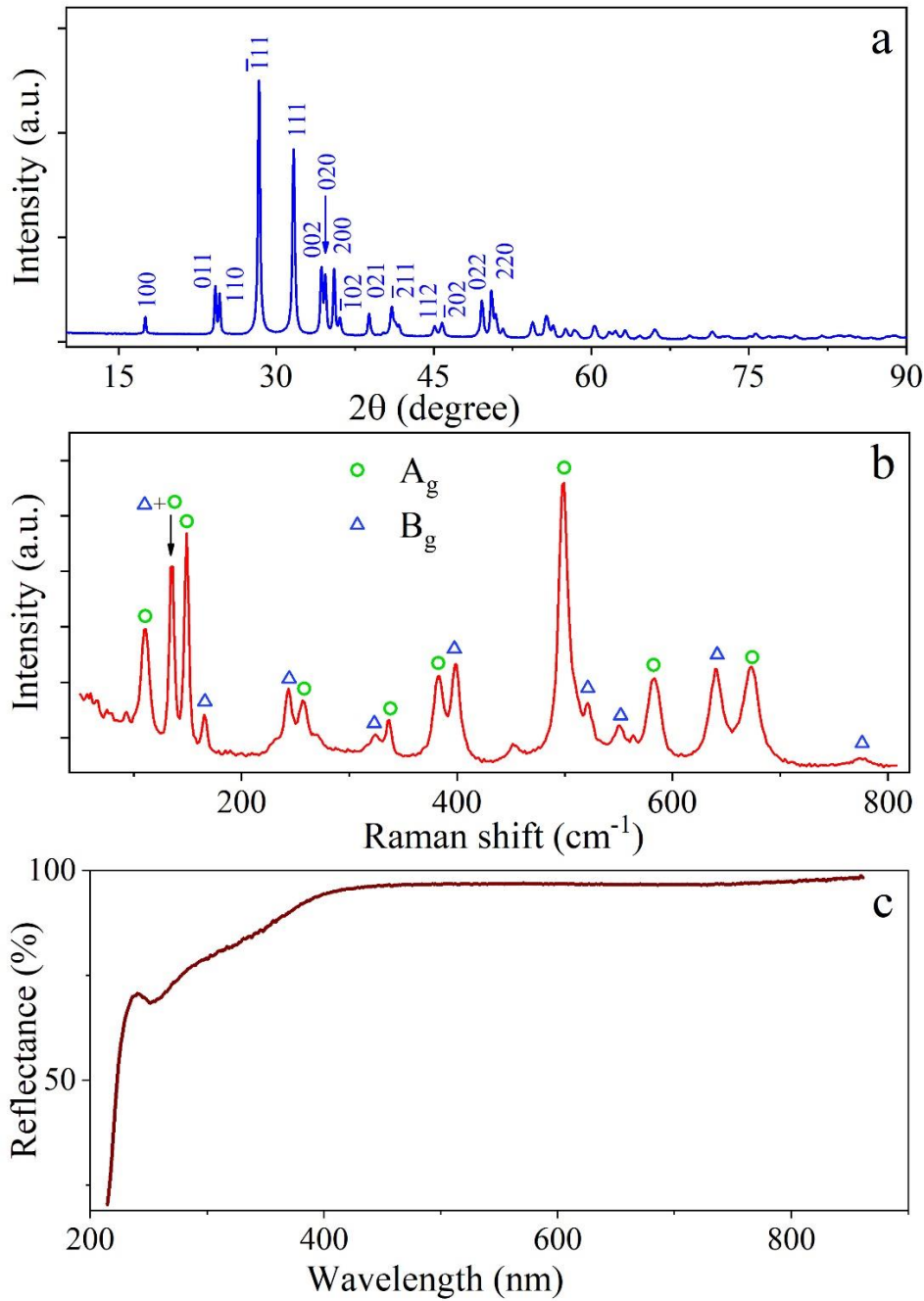

**FIGURE 2.** Characterization of the powder studied:
a) XRD, the diffractogram shows the Miller indices for $HfO_2$ with monoclinic crystal structure in accordance with [33];
b) Raman spectrum. Active vibrational modes are indicated in accordance with [34];
c) diffuse reflectance spectrum.

In Fig. 3 PL emission and PL excitation (PLE) spectra measured at different temperatures are presented. Top plot shows PL spectra measured at room temperature: the luminescence band with 2.6 eV maximum appears under UV excitation. The 4.2 eV emission, presented in the bottom graphs, is most efficiently excited by photons with an energy of 5.8 eV at temperatures < 200 K. It is seen that temperature decrease from 200 to 90 K leads to an increase in the PL intensity while further cooling to the 40 K is followed by slight decrease of the PL emission. In addition, with decreasing temperature, the studied peak becomes narrower and the maximum shifts slightly toward higher energies, from 4.08 eV at 205 K to 4.24 eV at 40 K. In turn, the position of the maximum for the excitation spectra remains almost unchanged in the investigated temperature range.

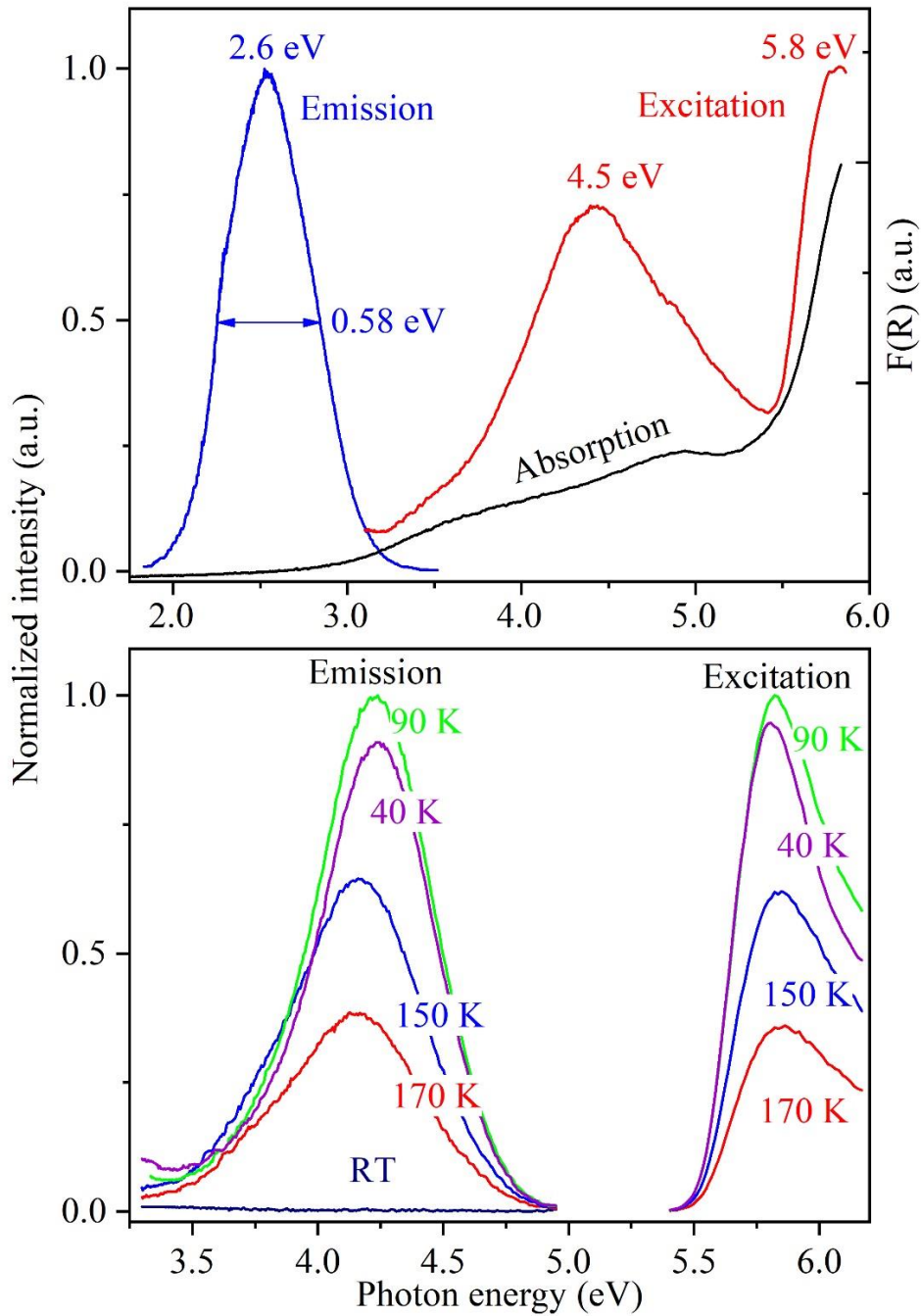

**FIGURE 3.** Hafnia PL, PLE and OA spectra measured at different temperatures. The top panel shows the spectra at room temperature; the bottom panel shows the spectra under study at different temperatures.

## 4. DISCUSSION

### *4.1 Estimation of the bandgap*

We have analyzed the DRS measured at room temperature using the Kubelka-Munk function *F(R)* [35]:

$$\alpha \sim F(R) = \frac{(1-R)^2}{2R} \quad (1)$$

where $\alpha$ is the absorption index, cm$^{-1}$. The resulting optical absorption (OA) spectrum is shown in Figure 3 (a top panel, a black line). It should be underscored that an increase in absorption at photon energies of $h\nu > 3.5$ eV, as well as the presence of a weakly pronounced shoulder in the 4.9 – 5.1 eV region is consistent with the presence of a 4.5 eV band in the photoluminescence excitation spectrum. When exposed to UV radiation, the sample exhibits a 2.6 eV emission associated with oxygen-deficient centers [3,36]. According to calculations [37,38], oxygen vacancies in positions of $O_{3f}$ and $O_{4f}$ in different charge states create additional energy levels in the forbidden gap, and corresponding optical transitions involve in forming spectral features of the OA spectrum. At $h\nu > 5.5$ eV, a sharp increase in absorption corresponds to the region of the intrinsic absorption edge, which is typical for the monoclinic hafnium dioxide [5,9,13,17,20,23,28,29,39] and quite coincides with the excitation energy of STE (see Figures 3 and 4).

To estimate the bandgap $E_g$, we have exploited the Tauc approach for various types of optical transitions [40]:

$$\alpha h\nu = B(h\nu - E_g)^n, \quad (2)$$

where *B* is a dimensional constant, *n* is an exponent depending on the type of optical transition.

Figure 4 outlines the calculated curves in Tauc coordinates for $n = 1/2$ (red circles) and $n = 2$ (blue squares). They characterize direct and indirect allowed band-to-band transitions, respectively. The widths of the direct (5.61 ± 0.05 eV) and indirect (5.31 ± 0.05 eV) bandgaps can be estimated by extrapolating the linear segments of the obtained curves to the intersection with the abscissa axis. It is worth stressing that the indirect allowed transitions dominate in the region of $h\nu < 5.68$ eV (highlighted in blue), whereas in the range of 5.68 eV $< h\nu <$ 5.76 eV (highlighted in violet) both the direct and indirect transitions form the OA spectrum within the region of the intrinsic edge. Finally, only the direct allowed transitions take place in the range of $h\nu > 5.76$ eV (highlighted in red). The inset in Figure 4 presents the OA spectrum at hand in ordinary coordinates. The spectrum approximation is shown over the entire range of photon energies, accounting for the dominance of one or another type of optical transitions:

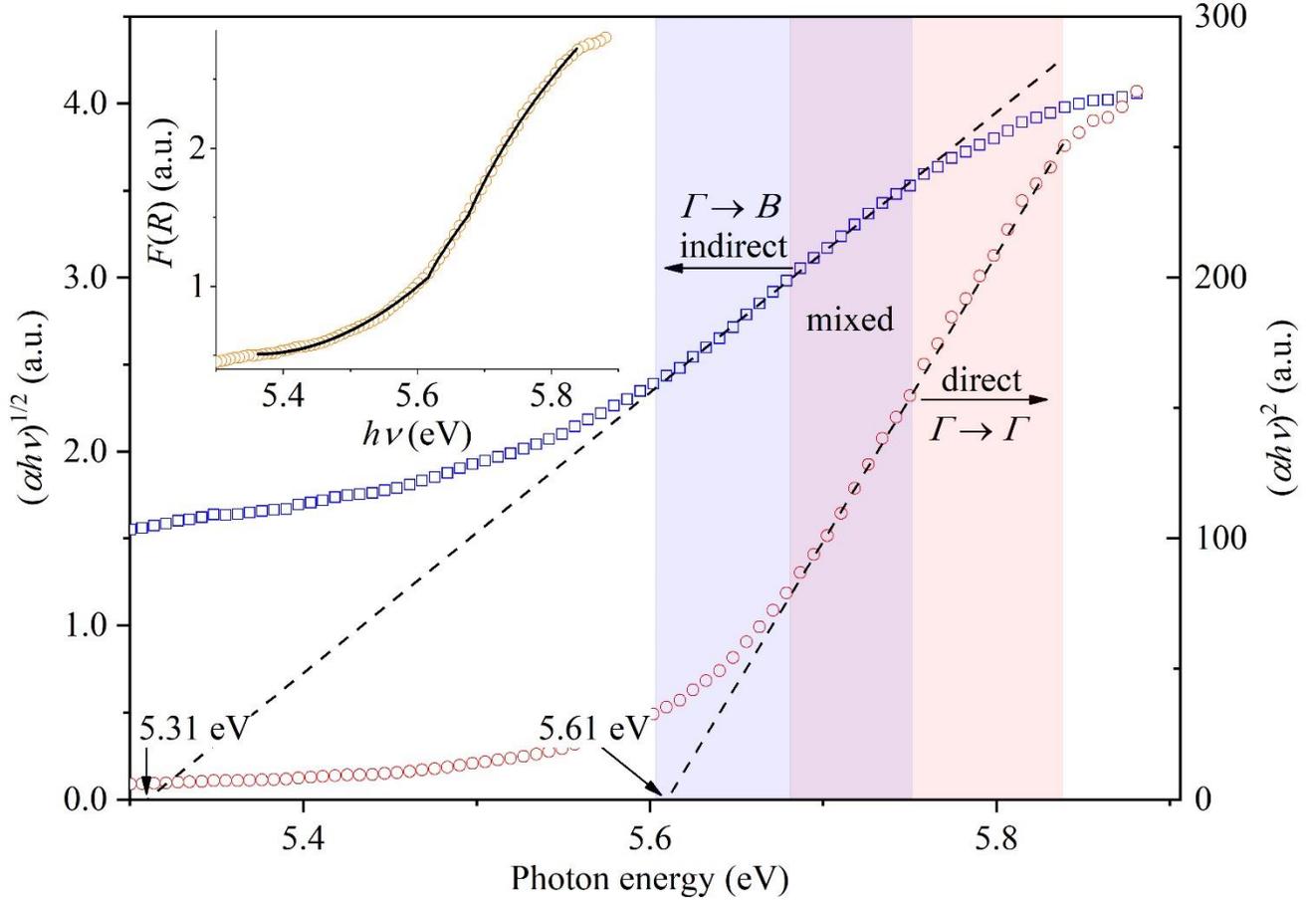

**FIGURE 4.** Determination of the bandgap width in m-HfO$_2$ powder studied. Blue squares indicate the Tauc plot for indirect band-to-band transitions. Red circles designate the Tauc plot for direct transitions. The energy ranges corresponding to the dominance of various types of transitions are highlighted in different colors (see the text in detail). In the inset, orange circles show an OA spectrum calculated through expression (1), the solid black line is the approximation using relations (3).

$$\alpha = \begin{cases} \dfrac{A_i}{h\upsilon}\left(h\upsilon - E_{gi}\right)^2, \ h\upsilon < E_1 & \text{indirect transitions dominate} \\[6pt] \dfrac{B_i}{h\upsilon}\left(h\upsilon - E_{gi}\right)^2 + \dfrac{B_d}{h\upsilon}\left(h\upsilon - E_{gd}\right)^{1/2}, \ E_1 < h\upsilon < E_2 & \text{both types of transitions occur} \\[6pt] \dfrac{C_d}{h\upsilon}\left(h\upsilon - E_{gd}\right)^{1/2}, \ h\upsilon > E_2 & \text{direct transitions dominate} \end{cases} \quad (3)$$

where $A_i$, $B_i$, $B_d$, $C_d$ are the corresponding dimensional constants, $E_1$ and $E_2$ are the boundaries of the energy ranges where various types of transitions occur, and the corresponding expressions are applicable.

The values of $E_1 = 5.61$ eV and $E_2 = 5.68$ eV are obtained from the above calculations. In addition, the approximation yields the width of the indirect ($E_{gi} = 5.36 \pm 0.05$ eV) and direct ($E_{gd} = 5.61 \pm 0.05$ eV) bandgaps. It should be emphasized that all the quantitative estimates are in excellent agreement with the Tauc plot data.

Table 1 provides independent data for the bandgap in m-HfO$_2$ structures with different morphologies to compare. Our estimates are in good agreement with these data. It should be noted that the monoclinic phase in the above structures of hafnium dioxide is dominant [9,17,23,28,39]. The papers [5,17,23] calculate the bandgap width in m-HfO$_2$ under the assumption of indirect optical transitions. Such an approach enables one to compute the value with an accuracy to the energy $\hbar\omega$ of phonons involved in three-particle band-to-band processes. In [28,29,39] the estimate is performed under the assumption of direct transitions. Thus, it can be inferred that the intrinsic absorption edge in nominally pure hafnium dioxide is formed by direct and indirect allowed band-to-band transitions alike. The possibility of simultaneous realization of two types of optical transitions has been previously revealed in thin films [9,20] and m-HfO$_2$ single crystals [13]. According to [41,42] an indirect allowed transition occurs between the points $\Gamma \to B$ of the Brillouin zone, and the direct one corresponds to the process $\Gamma \to \Gamma$.

**TABLE 1**. Estimates of a bandgap width in monoclinic HfO$_2$

| $E_g$, eV | | Morphology | Source |
|---|---|---|---|
| **indirect** | **direct** | | |
| $5.31 \pm 0.05$ | $5.61 \pm 0.05$ | powder | this work |
| 5.26 – 5.35 | — | thin film | [23] |
| 5.5 | — | thin film | [17] |
| 5.75 | — | nanoparticles | [5] |
| — | 5.3 – 5.8 | thin film | [28] |
| — | 5.57 – 5.62 | thin film | [29] |
| — | 5.4 – 5.7 | nanoparticles | [39] |
| 5.55 | 5.68 – 5.72 | thin film | [9] |
| 5.52 – 5.64 | 6.08 – 6.14 | thin film | [20] |
| 5.54 | 5.89 | single crystal | [13] |

### 4.2. *Spectral parameters of photoluminescence*

All the measured emission spectra were approximated with high accuracy using a broad single Gaussian-shaped band. The examples of the approximation for different temperatures can be seen in Figure 5. It should be noted that the observed PL bands at low intensities, characterized by a more extended low-energy wing (see the inset in Figure 5).

The visible deviation from approximation curve may be due to oxygen vacancies in various charge state. The vacancies create additional energy levels in the forbidden gap, which are 2.6 – 3.5 eV above the top of the valence band [37,38]

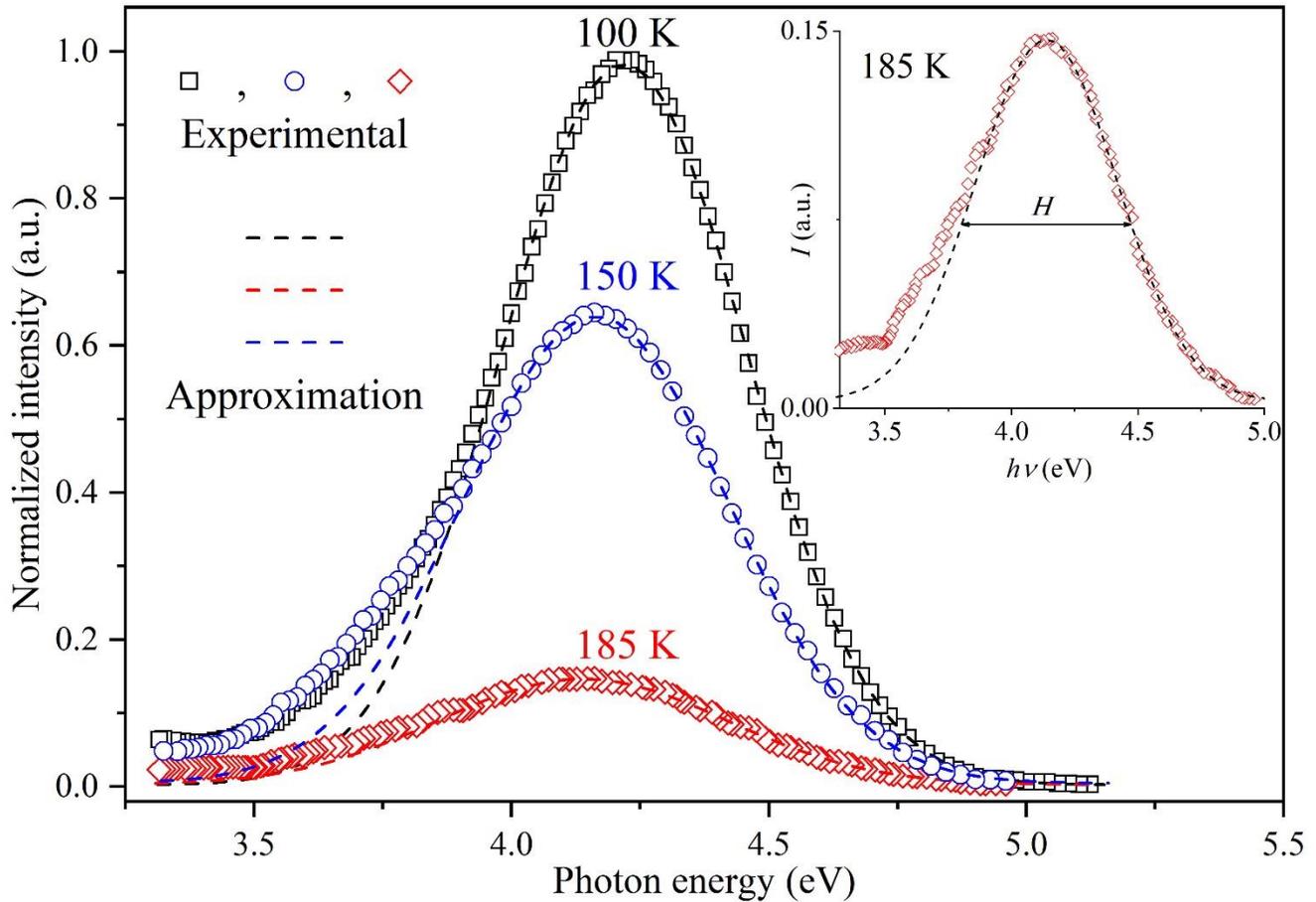

**FIGURE 5.** Hafnium dioxide PL spectra measured at different temperatures and approximated by a Gaussian-shaped peak. The inset depicts the results of the approximation of the emission spectrum at 185 K and the estimates of the half-width of the band.

### 4.2.1 *Thermal quenching of photoluminescence*

Figure 6 shows the experimental temperature dependence $I(T)$ of the PL intensity of the m-$HfO_2$ powder. Attention may be drawn to the fact that, in the temperature range of 40-90 K, the emission intensity increases and reaches a maximum at 90 K. At $T > 90$ K, a decline in intensity takes place, and at 200 K no emission at hand is registered. Thus, there are two regions with different behavior of the $I(T)$ dependence: enhancement (negative quenching) at $T < 90$ K and quenching at $T > 90$ K. Analysis of the temperature dependence $I(T)$ of the luminescence intensity was carried out applying the modified Mott relation [43,44]:

$$I(T) = I_0 \frac{1 + P_{NQ} \exp\left(-\frac{E_{NQ}}{kT}\right)}{1 + P_Q \exp\left(-\frac{E_Q}{kT}\right)} \quad (4)$$

where $I_0$ is the emission intensity at $T \to 0$ K; $P_{NQ}$ and $P_Q$ are corresponding pre-exponential factor; $E_{NQ}$ is the activation energy of the emission enhancement process, eV; $E_Q$ is the activation energy of the emission enhancement process, eV; $k$ is the Boltzmann constant, eV · K$^{-1}$.

Figure 6 depicts the experimental data of $I(T)$ approximated by the expression (4). For comparison, the values of the parameters calculated during the approximation and the known independent data are given in Table 2. The inset in Figure 6 includes the plot in Arrhenius coordinates. The linear segment ($R^2 = 0.97$) for $T > 90$ K points to the activation origin of the PL temperature quenching.

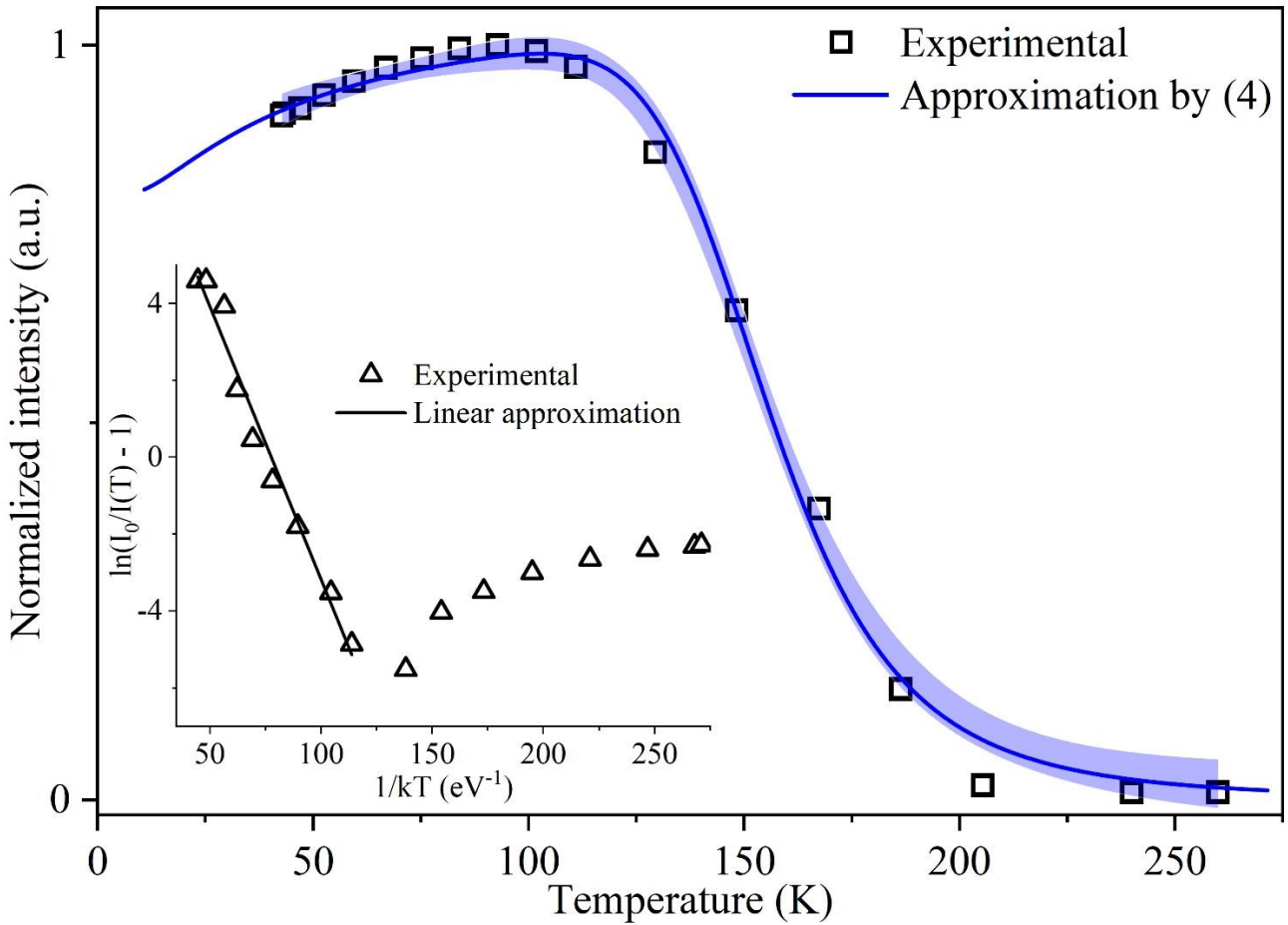

**FIGURE 6.** Temperature dependence of the emission intensity for the 4.2 eV band. The experimental values (square symbols) are approximated using the modified Mott relation (4) (blue line), blue shaded area along the line indicates 95% confidence interval. The inset contains the data for the temperature quenching of the investigated luminescence plotted in Arrhenius coordinates and their linear approximation (black line) in the corresponding range.

**TABLE 2** Parameters calculated upon approximating the PL temperature dependencies in comparison with the independent data for m-HfO$_2$.

| Parameter | Value | Source |
|---|---|---|
| $P_Q$, ×10$^5$ | 0.3 – 0.8 <br> 0.3 – 3.2 | eq. (4) <br> [10] |
| $P_{NQ}$ | 0.35 ± 0.05 | eq. (4) |
| $E_Q$, meV | 140 ± 20 <br> 180 – 230 | eq. (4) <br> [10] |
| $E_{NQ}$, meV | 3 ± 0.5 | eq. (4) |
| $E_m(0)$, eV | 4.24 <br> 4.40*(10 K) <br> 4.26*(10 K) <br> 4.30*(20 K) | eq. (5) <br> [9] <br> [30] <br> [10] |
| $A_s$, eV | 0.77 ± 0.23 | eq. (5) |
| $\hbar\omega_s$, meV | 32 ± 5 | eq. (5) |
| $H(0)$, eV | 0.54 ± 0.01 <br> 0.51*(10 K) <br> 0.72*(10 K) <br> 0.80*(20 K) | eq. (6) <br> [9] <br> [30] <br> [10] |
| $A_b$, eV | 1.0 ± 0.2 | eq. (6) |
| $\hbar\omega_b$, meV | 36 ± 5 | eq. (6) |
| $S$ | 12.0 <br> 24.0 <br> 40.6 | eq. (8a) <br> eq. (8b) <br> eq. (8c) |
| $D_s/D_b$ | 1.07 | eq. (7) |

\* estimates based on the data presented in the reference

The energy $E_Q$ could control the magnitude of the barrier for a thermally activated non-radiative transition from the luminescent state of a self-trapped exciton and an F-center – hole pair on the same adiabatic potential surface [45]. The calculated value of the energy $E_{NQ}$ proves to be close to that of the thermally activated barrier to be overcome for self-trapping of a free exciton in m-HfO$_2$ [46]. For example, in alkali halide crystals, luminescence of free excitons is observed at temperatures of < 40 K, and the thermal barrier for self-trapping resides in the range of 11–33 meV [46]. The thermal quenching parameters computed are nearby the radioluminescence data for HfO$_2$ nanocrystals with a monoclinic structure [10], see Table 2.

### 4.2.2 *Temperature dependencies of spectral characteristics*

The temperature dependence of the position of the maximum $E_m(T)$ can be described using Fan expression [47,48] for the displacement of energy levels in solids:

$$E_m(T) = E_m(0) - A_s \langle n_s \rangle \quad (5)$$

where $E_m(0)$ is the position of the maximum at zero temperature, eV; $A_s$ is the Fan parameter depending on the microscopic properties of the material and is related to the second-order deformation potential constant, eV [47]; $\langle n_s \rangle = (\exp(\hbar\omega_s / kT) - 1)^{-1}$ is the Bose-Einstein factor for phonons with energy $\hbar\omega_s$, responsible for the shift of electronic levels.

In Figure 7, the red solid line denotes the approximation by using the expression (5). The values of the parameter calculated are presented in Table 2. It should be noted the maximum energy $E_m(0)$ is close to the experimentally values measured at 10 K [30] and 20 K [10].

The observable temperature-change in the half-width $H(T)$ of the luminescence band was described using the following expression [49,50] (blue line in Figure 7):

$$H(T) = H(0) + A_b \langle n_b \rangle \quad (6)$$

where $H(0)$ is the half-width at zero temperature, eV; $\langle n_b \rangle = (\exp(\hbar\omega_b / kT) - 1)^{-1}$ is the Bose-Einstein factor for phonons with energy $\hbar\omega_b$, responsible for the broadening of energy levels; $A_b$ is the approximation parameter specified by the exciton-phonon interaction, eV. In [49] the expression (6) contains an additional linearly temperature-dependent summand that accounts for the interaction between excitons and acoustic phonons at high temperatures. When approximating the experimental data, the foregoing summand was neglected due to its smallness. The value of $H(0)$ is in a good agreement with the experimentally measured half-width of the 4.4 eV band at 10 K in [9]. However, the former is noticeably lower in comparison with the corresponding values of the 4.3 eV band at 10 K [30] and 20 K [10].

The found shift of maximum position and band narrowing with decreasing temperature are a typical behavior when the phonon-induced displacement of energy levels occurs [51,52]. As can be inferred from approximating the temperature dependencies, phonons with close energies $\hbar\omega_s = 32$ meV and $\hbar\omega_b = 36$ meV are responsible for both the shift and the broadening of the electron levels. The predicted energies of effective phonons match the maximum in the m-HfO$_2$ IR spectrum being near 300 cm$^{-1}$ (37 meV) and caused by one longitudinal $A_u$ and two transverse $B_u$ vibrational modes assigned to oxygen atoms movement [53].

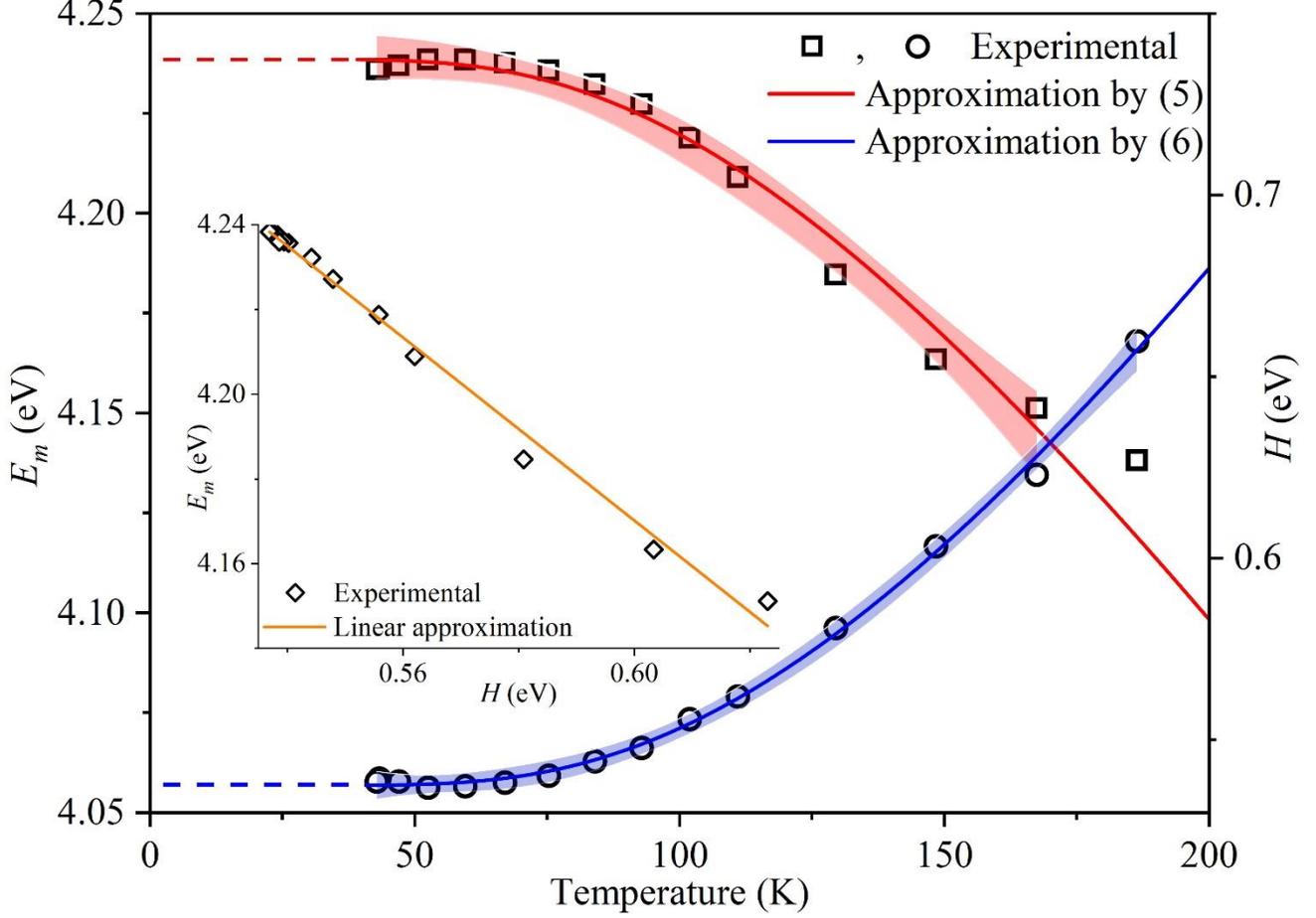

**FIGURE 7.** Temperature dependencies of spectral characteristics for the STE emission band. The squares show the positions of the maximum; the circles indicate the half-width of the band. The red and blue lines mark the appropriate approximations using the relations (5) and (6), blue and red shaded areas along the lines indicate 99% confidence intervals. The orange line in the inset is the linear approximation of the $E_m(H)$ experimental dependence.

The temperature functions $E_m(T)$ and $H(T)$ are related to each other by the equation [52]:

$$E_m(T) = E_m' - \frac{D_s}{D_b} H(T). \tag{7}$$

Here $E_m'$ – is the virtual position of the maximum of the band at zero peak half-width. $E_m'$ differs from $E_m(0)$ due to zero "in-frozen" vibrations [54]. $D_s$ и $D_b$ – are the corresponding constants of the second-order dynamic deformation potential. They abide by the processes of displacement and broadening of energy levels [52].

The inset to Figure 7 shows the $E_m(H)$ linearly approximated experimental dependence. The estimate of the $D_s/D_b$ ratio was performed along the slope of the calculated straight line (see the value in Table 2).

The one-phonon model used, and the Gaussian shape of the peak studied allows one to estimate the Huang-Rhys factor $S$ that characterizes the degree of electron-phonon interaction through the following expressions [51,55]:

$$A_s = 2S\hbar\omega_s \qquad (8a)$$

$$\Delta E = \hbar\omega_s(2S+1) \qquad (8b)$$

$$S = \frac{H^2(0)}{8\ln 2(\hbar\omega_b)^2} \qquad (8c)$$

where $\Delta E$ is Stokes shift, eV.

The calculated values of $S$ are listed in Table 2. The expression (8b)-based computation was carried out for a temperature of 40 K, when $\Delta E = 1.57$ eV.

Unfortunately, we failed to collect independent quantitative data on the characteristics of the electron-phonon interaction in m-HfO$_2$. However, the obtained values of $S \gg 1$ and the estimates of the phonon energy enable making a conclusion that a strong exciton-phonon interaction holds in the HfO$_2$ powder studied. High magnitudes of the Stokes shift and a large half-width indicate that the processes of STE radiative decay underlie the mechanisms of forming the 4.2 eV emission [46].

The predominantly ionic type of bond is inherent to m-HfO$_2$ perfect crystals, and the corresponding theoretical predictions of [56] show that self-trapping of charge carriers and the emergence of STE are possible. In particular, holes are localized near the top of the valence band formed by oxygen 2p-orbitals [56]. The localization occurs due to insignificant displacements of non-equivalent O$_{3f}$ and O$_{4f}$ atoms. In this case, as noted above, STE intensively interacts with the $A_u$ and $B_u$ oxygen vibrational modes. The oxygen vacancies can enhance this effect including once changed their charge state. In the process, there arises significant (up to 11% for a fully ionized vacancy) distortion of the Hf–O bond and the creation of an appropriate potential well for a hole [37,38]. In addition, the mutually approaching atoms cause an increase in the covalent contribution to this bond and provoke the localization of an electron [45]. Note that the capture of an electron by an oxygen vacancy is energetically favorable [45]. This fact manifests itself in the coexistence of excitons and optically active F-type centers in different charge states. This is in good agreement with our findings, since with band-to-band excitation, a 2.6 eV blue luminescence typical for oxygen vacancies is observed simultaneously with the STE emission studied (see Figure 3). Relying on the analysis of the decay kinetics, we have ascribed the emission mentioned above to F and F$^+$-centers in HfO$_2$, as a result of the processes of their excitation and charge redistribution [30].

# CONCLUSION

This work is aimed at investigating the diffuse reflectance of a nanostructured powder of monoclinic hafnium dioxide at room temperature and its photoluminescent properties in the range of 40 – 300 K. The DRS analysis was carried out within the Kubelka-Munk formalism. The spectral peculiarities of the obtained OA spectra in the region of 3.5 – 5.1 eV point to the presence of oxygen-deficient centers in the hafnia at hand. When studying the intrinsic absorption edge through the Tauc plot, it is found that allowed optical transitions of two types are realized. We have calculated the bandgap $E_g$ under the assumption made for indirect (5.31 ± 0.05eV) and direct (5.61 ± 0.05eV) allowed transitions and run a comparative analysis of the obtained values with independent data.

At $T < 200$ K, the PL spectra of the m-$HfO_2$ powder contain band with a maximum position of 4.2 eV. The temperature behavior of the spectral characteristics and the magnitude of the Huang-Rhys factor $S = 12.0 – 40.6$ evidence STE radiative decay processes underlying the mechanisms of forming the indicated emission. In this work, for the first time, we have estimated the energies of effective phonons $\hbar\omega_s = 32$ meV and $\hbar\omega_b = 36$ meV, responsible for the shift and broadening of energy levels, respectively. The calculated vibrational energies are close to the predicted maximum in the IR spectrum, which is due to an $A_u+2B_u$ superposition of longitudinal and transverse vibrational modes of oxygen atoms. In this regard, it can be concluded that holes having localized near the top of the valence band formed by the electron shells of oxygen atoms are the reason for STE in m-$HfO_2$.

The obtained temperature behavior for the 4.2 eV emission was analyzed using the modified Mott relation. In addition, the activation energies of quenching ($E_Q = 140$ meV) and enhancement ($E_{NQ} = 3$ meV) of the STE luminescence were calculated. It is shown that the magnitude of $E_{NQ}$ can be due to the height of a barrier needed to overcome for a free exciton to be self-trapped. Thorough study of the discussed excitonic effects can advance development of hafnia-based structures with controlled optical response of intrinsic origin.

# ACKNOWLEDGMENTS

The work was supported by Minobrnauki research project FEUZ-2020-0059.The work was supported by Minobrnauki research project FEUZ-2020-0059.

# REFERENCES

[1] J.T. Gaskins, P.E. Hopkins, D.R. Merrill, S.R. Bauers, E. Hadland, D.C. Johnson, D. Koh, J.H. Yum, S. Banerjee, B.J. Nordell, M.M. Paquette, A.N. Caruso, W.A. Lanford, P. Henry, L. Ross, H. Li, L. Li, M. French, A.M. Rudolph, S.W. King, Review — Investigation and review of the thermal, mechanical, electrical, optical, and structural properties of atomic layer deposited high-k dielectrics: Beryllium oxide, aluminum oxide, hafnium oxide, and aluminum nitride, ECS J. Solid State Sci. Technol. 6 (2017) N189–N208. https://doi.org/10.1149/2.0091710jss.

[2] C. Zhao, J. Xiang, Atomic layer deposition (ALD) of metal gates for CMOS, Appl. Sci. 9 (2019). https://doi.org/10.3390/app9112388.

[3] V.A. Gritsenko, D.R. Islamov, T. V. Perevalov, V.S. Aliev, A.P. Yelisseyev, E.E. Lomonova, V.A. Pustovarov, A. Chin, Oxygen Vacancy in Hafnia as a Blue Luminescence Center and a Trap of Charge Carriers, J. Phys. Chem. C. 120 (2016) 19980–19986. https://doi.org/10.1021/acs.jpcc.6b05457.

[4] M. Kirm, J. Aarik, M. Jürgens, I. Sildos, Thin films of HfO 2 and ZrO 2 as potential scintillators, Nucl. Instruments Methods Phys. Res. Sect. A Accel. Spectrometers, Detect. Assoc. Equip. 537 (2005) 251–255. https://doi.org/10.1016/j.nima.2004.08.020.

[5] J. Manikantan, H.B. Ramalingam, B.C. Shekar, B. Murugan, R.R. Kumar, J.S. Santhoshi, Physical and optical properties of HfO2 NPs – Synthesis and characterization in finding its feasibility in opto-electronic devices, Adv. Powder Technol. 28 (2017) 1636–1646. https://doi.org/10.1016/j.apt.2017.03.022.

[6] H. Mulaosmanovic, E.T. Breyer, S. Dünkel, S. Beyer, T. Mikolajick, S. Slesazeck, Ferroelectric field-effect transistors based on HfO2: A review, Nanotechnology. 32 (2021). https://doi.org/10.1088/1361-6528/ac189f.

[7] W.T. Fan, P.T. Liu, P.Y. Kuo, C.M. Chang, I.H. Liu, Y. Kuo, Numerical analysis of oxygen-related defects in amorphous in-w-o nanosheet thin-film transistor, Nanomaterials. 11 (2021) 3070. https://doi.org/10.3390/nano11113070.

[8] I. Zrinski, C.C. Mardare, L.I. Jinga, J.P. Kollender, G. Socol, A. Minenkov, A.W. Hassel, A.I. Mardare, Electrolyte-dependent modification of resistive switching in anodic hafnia, Nanomaterials. 11 (2021) 1–18. https://doi.org/10.3390/nano11030666.

[9] J. Aarik, H. Mändar, M. Kirm, L. Pung, Optical characterization of HfO2 thin films grown by atomic layer deposition, Thin Solid Films. 466 (2004) 41–47. https://doi.org/10.1016/j.tsf.2004.01.110.

[10] I. Villa, A. Lauria, F. Moretti, M. Fasoli, C. Dujardin, M. Niederberger, A. Vedda, Radio-luminescence spectral features and fast emission in hafnium dioxide nanocrystals, Phys. Chem. Chem. Phys. 20 (2018) 15907–15915. https://doi.org/10.1039/c8cp01230j.

[11] H. Arashi, Pressure-Induced Phase Transformation of HfO2, J. Am. Ceram. Soc. 75 (1992) 844–847. https://doi.org/10.1111/j.1151-2916.1992.tb04149.x.

[12] A. Ortega, E.J. Rubio, K. Abhilash, C. V. Ramana, Correlation between phase and optical properties of yttrium-doped hafnium oxide nanocrystalline thin films, Opt. Mater. (Amst). 35 (2013) 1728–1734. https://doi.org/10.1016/j.optmat.2013.05.017.

[13] V. V. Lozanov, N.I. Baklanova, V.R. Shayapov, A.S. Berezin, Crystal Growth and Photoluminescence Properties of Reactive CVD-Derived Monoclinic Hafnium Dioxide, Cryst. Growth Des. 16 (2016) 5283–5293. https://doi.org/10.1021/acs.cgd.6b00824.

[14] A.O. Shilov, S.S. Savchenko, A.S. Vokhmintsev, A. V. Chukin, M.S.

Karabanalov, M.I. Vlasov, I.A. Weinstein, Energy gap evaluation in microcrystalline m-hfo2 powder, J. Sib. Fed. Univ. - Math. Phys. 14 (2021) 224–229. https://doi.org/10.17516/1997-1397-2021-14-2-224-229.

[15] B.K. Kim, H.O. Hamaguchi, Raman spectrum of 18O-labelled hafnia, Mater. Res. Bull. 32 (1997) 1367–1370. https://doi.org/10.1016/S0025-5408(97)00114-1.

[16] D.A. Pejaković, Studies of the phosphorescence of polycrystalline hafnia, J. Lumin. 130 (2010) 1048–1054. https://doi.org/10.1016/j.jlumin.2010.01.023.

[17] M. Kong, B. Li, C. Guo, P. Zeng, M. Wei, W. He, The optical absorption and photoluminescence characteristics of evaporated and IAD HfO 2 thin films, Coatings. 9 (2019). https://doi.org/10.3390/coatings9050307.

[18] A. Vinod, M.S. Rathore, N. Srinivasa Rao, Effects of annealing on quality and stoichiometry of HfO2 thin films grown by RF magnetron sputtering, Vacuum. 155 (2018) 339–344. https://doi.org/10.1016/j.vacuum.2018.06.037.

[19] M. Dhanunjaya, S.A. Khan, A.P. Pathak, D.K. Avasthi, S.V.S. Nageswara Rao, Ion induced crystallization and grain growth of hafnium oxide nano-particles in thin-films deposited by radio frequency magnetron sputtering, J. Phys. D. Appl. Phys. 50 (2017). https://doi.org/10.1088/1361-6463/aa9723.

[20] X. Fan, H. Liu, B. Zhong, C. Fei, X. Wang, Q. Wang, Optical characteristics of H2O-based and O3-based HfO2 films deposited by ALD using spectroscopy ellipsometry, Appl. Phys. A Mater. Sci. Process. 119 (2015) 957–963. https://doi.org/10.1007/s00339-015-9048-9.

[21] R. Kumar, V. Chauhan, N. Koratkar, S. Kumar, A. Sharma, K.H. Chae, S.O. Won, Influence of high energy ion irradiation on structural, morphological and optical properties of high-k dielectric hafnium oxide (HfO2) thin films grown by atomic layer deposition, J. Alloys Compd. 831 (2020) 154698. https://doi.org/10.1016/j.jallcom.2020.154698.

[22] A.O. Shilov, S.S. Savchenko, A.S. Vokhmintsev, V.A. Gritsenko, I.A. Weinstein, Optical parameters and energy gap estimation in Hafnia Thin film, AIP Conf. Proc. 2313 (2020). https://doi.org/10.1063/5.0033176.

[23] X. Luo, Y. Li, H. Yang, Y. Liang, K. He, W. Sun, H.H. Lin, S. Yao, X. Lu, L. Wan, Z. Feng, Investigation of HfO2 thin films on Si by X-ray photoelectron spectroscopy, rutherford backscattering, grazing incidence X-ray diffraction and Variable Angle Spectroscopic Ellipsometry, Crystals. 8 (2018) 1–16. https://doi.org/10.3390/cryst8060248.

[24] V.A. Pustovarov, T.P. Smirnova, M.S. Lebedev, V.A. Gritsenko, M. Kirm, Intrinsic and defect related luminescence in double oxide films of Al-Hf-O system under soft X-ray and VUV excitation, J. Lumin. 170 (2016) 161–167. https://doi.org/10.1016/j.jlumin.2015.10.053.

[25] S. Chen, Z. Liu, L. Feng, X. Zhao, Photoluminescent properties of undoped and Ce-doped HfO2 thin films prepared by magnetron sputtering, J. Lumin. 153 (2014) 148–151. https://doi.org/10.1016/j.jlumin.2014.03.017.

[26] K. Fiaczyk, A.J. Wojtowicz, E. Zych, Photoluminescent properties of monoclinic

HfO2:Ti sintered ceramics in 16-300 K, J. Phys. Chem. C. 119 (2015) 5026–5032. https://doi.org/10.1021/jp512685u.

[27] A. Ćirić, S. Stojadinović, M.D. Dramićanin, Judd-Ofelt and chromaticity analysis of hafnia doped with trivalent europium as a potential white LED phosphor, Opt. Mater. (Amst). 88 (2019) 392–395. https://doi.org/10.1016/j.optmat.2018.11.056.

[28] M.C. Cheynet, S. Pokrant, F.D. Tichelaar, J.L. Rouvìre, Crystal structure and band gap determination of Hf O2 thin films, J. Appl. Phys. 101 (2007). https://doi.org/10.1063/1.2697551.

[29] D. Franta, I. Ohlídal, D. Nečas, F. Vižd'A, O. Caha, M. Hasoň, P. Pokorný, Optical characterization of HfO2 thin films, Thin Solid Films. 519 (2011) 6085–6091. https://doi.org/10.1016/j.tsf.2011.03.128.

[30] E. Aleksanyan, M. Kirm, E. Feldbach, V. Harutyunyan, Identification of F+ centers in hafnia and zirconia nanopowders, Radiat. Meas. 90 (2016) 84–89. https://doi.org/10.1016/j.radmeas.2016.01.001.

[31] W.H. Hall, G.K. Williamson, The diffraction pattern of cold worked metals: I The nature of extinction, Proc. Phys. Soc. Sect. B. 64 (1951) 937–946. https://doi.org/10.1088/0370-1301/64/11/301.

[32] W.H. Hall, X-ray line broadening in metals [3], Proc. Phys. Soc. A. 62 (1949) 741–743. https://doi.org/10.1088/0370-1298/62/11/110.

[33] D.M. Adams, S. Leonard, D.R. Russell, R.J. Cernik, X-ray diffraction study of Hafnia under high pressure using synchrotron radiation, J. Phys. Chem. Solids. 52 (1991). https://doi.org/10.1016/0022-3697(91)90052-2.

[34] R. Wu, B. Zhou, Q. Li, Z. Jiang, W. Wang, W. Ma, X. Zhang, Elastic and vibrational properties of monoclinic HfO 2 from first-principles study, J. Phys. D. Appl. Phys. 45 (2012). https://doi.org/10.1088/0022-3727/45/12/125304.

[35] P. Kubelka, F. Munk, Ein Beitrag zur Optik der Farbanstriche, Zeitschrift Für Tech. Phys. 12 (1931).

[36] T. V. Perevalov, V.S. Aliev, V.A. Gritsenko, A.A. Saraev, V. V. Kaichev, Electronic structure of oxygen vacancies in hafnium oxide, Microelectron. Eng. 109 (2013). https://doi.org/10.1016/j.mee.2013.03.005.

[37] D. Muñoz Ramo, J.L. Gavartin, A.L. Shluger, G. Bersuker, Spectroscopic properties of oxygen vacancies in monoclinic Hf O2 calculated with periodic and embedded cluster density functional theory, Phys. Rev. B - Condens. Matter Mater. Phys. 75 (2007) 1–12. https://doi.org/10.1103/PhysRevB.75.205336.

[38] J.L. Gavartin, D.M. Ramo, A.L. Shluger, G. Bersuker, B.H. Lee, Negative oxygen vacancies in HfO 2 as charge traps in high-k stacks, Appl. Phys. Lett. 89 (2006). https://doi.org/10.1063/1.2236466.

[39] N.G. Semaltianos, J.M. Friedt, R. Chassagnon, V. Moutarlier, V. Blondeau-Patissier, G. Combe, M. Assoul, G. Monteil, Oxide or carbide nanoparticles synthesized by laser ablation of a bulk Hf target in liquids and their structural, optical, and dielectric properties, J. Appl. Phys. 119 (2016). https://doi.org/10.1063/1.4951740.

[40] J. Tauc, Optical Properties of Amorphous Semiconductors, in: Amorph. Liq. Semicond., Springer US, 1974: pp. 159–220. https://doi.org/10.1007/978-1-4615-8705-7_4.

[41] Q.J. Liu, N.C. Zhang, F.S. Liu, Z.T. Liu, Structural, electronic, optical, elastic properties and Born effective charges of monoclinic HfO2 from first-principles calculations, Chinese Phys. B. 23 (2014). https://doi.org/10.1088/1674-1056/23/4/047101.

[42] T.V. Perevalov, V.A. Gritsenko, Application and electronic structure of high-permittivity dielectrics, Uspekhi Fiz. Nauk. 180 (2010) 587. https://doi.org/10.3367/ufnr.0180.201006b.0587.

[43] A.S. Vokhmintsev, I.A. Weinstein, Temperature effects in 3.9 eV photoluminescence of hexagonal boron nitride under band-to-band and subband excitation within 7–1100 K range, J. Lumin. 230 (2021) 117623. https://doi.org/10.1016/j.jlumin.2020.117623.

[44] H. Shibata, Negative thermal quenching curves in photoluminescence of solids, Japanese J. Appl. Physics, Part 1 Regul. Pap. Short Notes Rev. Pap. 37 (1998). https://doi.org/10.1143/jjap.37.550.

[45] R.T. Williams, K.S. Song, The self-trapped exciton, J. Phys. Chem. Solids. 51 (1990) 679–716. https://doi.org/10.1016/0022-3697(90)90144-5.

[46] I. Pelant, J. Valenta, Luminescence Spectroscopy of Semiconductors, 2012. https://doi.org/10.1093/acprof:oso/9780199588336.001.0001.

[47] H.Y. Fan, Temperature dependence of the energy gap in semiconductors, Phys. Rev. 82 (1951) 900–905. https://doi.org/10.1103/PhysRev.82.900.

[48] I.A. Vainshtein, A.F. Zatsepin, V.S. Kortov, Applicability of the empirical varshni relation for the temperature dependence of the width of the band gap, Phys. Solid State. 41 (1999) 905–908. https://doi.org/10.1134/1.1130901.

[49] S. Rudin, T.L. Reinecke, B. Segall, Temperature-dependent exciton linewidths in semiconductors, Phys. Rev. B. 42 (1990) 218–231.

[50] S.S. Savchenko, I.A. Weinstein, Inhomogeneous broadening of the exciton band in optical absorption spectra of InP/ZnS nanocrystals, Nanomaterials. 9 (2019). https://doi.org/10.3390/nano9050716.

[51] B.D. Evans, M. Stapelbroek, Optical properties of the F+ center in crystalline Al2O3, Phys. Rev. B. 18 (1978). https://doi.org/10.1103/PhysRevB.18.7089.

[52] I.A. Weinstein, V.S. Kortov, The shape and the temperature dependence of the main band in UV absorption spectra of TLD-500 dosimetric crystals, in: Radiat. Meas., 2001. https://doi.org/10.1016/S1350-4487(01)00102-0.

[53] B. Zhou, H. Shi, X.D. Zhang, Q. Su, Z.Y. Jiang, The simulated vibrational spectra of HfO2 polymorphs, J. Phys. D. Appl. Phys. 47 (2014). https://doi.org/10.1088/0022-3727/47/11/115502.

[54] I.A. Weinstein, A.F. Zatsepin, Modified Urbach's rule and frozen phonons in glasses, Phys. Status Solidi C Conf. 1 (2004) 2916–2919. https://doi.org/10.1002/pssc.200405416.

[55] K.P. O'Donnell, X. Chen, Temperature dependence of semiconductor band gaps, Appl. Phys. Lett. 58 (1991). https://doi.org/10.1063/1.104723.

[56] D. Muñoz Ramo, A.L. Shluger, J.L. Gavartin, G. Bersuker, Theoretical prediction of intrinsic self-trapping of electrons and holes in monoclinic HfO2, Phys. Rev. Lett. 99 (2007) 1–4. https://doi.org/10.1103/PhysRevLett.99.155504.